\documentclass[prl,twocolumn,superscriptaddress,amsmath,amssymb]{revtex4-2}
\usepackage{graphicx}
\usepackage{bm}
\usepackage[colorlinks=true,linkcolor=blue,citecolor=blue,urlcolor=blue]{hyperref}

\newcommand{\Qp}{\mathbb{Q}_p}
\newcommand{\Zp}{\mathbb{Z}_p}
\newcommand{\norm}[1]{\Vert #1\Vert_p}
\newcommand{\tphi}{\tilde\varphi}
\newcommand{\tpsi}{\tilde\psi}
\newcommand{\Dsig}{\boldsymbol{D}^{\sigma}}

\begin{document}

\title{Hierarchical connectivity sets the upper critical dimension of neural avalanches}

\author{W.~A. Z\'u\~niga-Galindo}
\email{wilson.zunigagalindo@utrgv.edu}
\affiliation{University of Texas Rio Grande Valley, School of Mathematical \& Statistical
Sciences, One West University Blvd, Brownsville, TX 78520, United States}

\date{Draft of \today}

\begin{abstract}
Neural avalanches with size exponent $3/2$ are usually explained by mean-field critical
branching, which for short-range coupling requires dimension $N\ge4$. We show that
hierarchical connectivity removes this constraint. In a $p$-adic version of the
Buice--Cowan field theory, a coupling that decays by $p^{-(N+\sigma)}$ per hierarchical
level selects the Taibleson--Vladimirov operator, and the critical theory is ultrametric
directed percolation with upper critical dimension $N_c=2\sigma$. Exact simulations on a
tree give $\tau_S=3/2$ for $\sigma\le N/2$, decreasing continuously for $\sigma>N/2$.
\end{abstract}

\maketitle

Cortical activity propagates in cascades, neuronal avalanches, whose sizes are
distributed as $S^{-3/2}$~\cite{BeggsPlenz2003}. The standard explanation is that the
cortex operates near the critical point of an absorbing-state transition, and that $3/2$
is the mean-field avalanche exponent of a critical branching process. Buice and Cowan made
this precise: their field theory of spiking networks maps onto Reggeon field theory, so
the transition is in the directed-percolation (DP) class~\cite{BuiceCowan2007}. This
raises a difficulty. For short-range coupling, DP has upper critical dimension $N_c=4$,
and below it the avalanche exponent differs from $3/2$; in one dimension,
$\tau_S\simeq1.11$~\cite{Munoz1999}. A cortical sheet is effectively two-dimensional (three at most, counting layers),
well below $N_c=4$; here, as in Ref.~\cite{BuiceCowan2007}, $N$ counts the dimensions of the
neuron space, and time is an additional coordinate. In a Euclidean theory with local
coupling, cortical avalanches would therefore not be mean-field. But cortical connectivity
is hierarchical. Sporns and co-workers proposed that the brain's structural connections
are arranged in self-similar, nested patterns across multiple spatial scales, from local
circuits to whole brain regions~\cite{Sporns2006,CostaSporns2005,SpornsTononiKotter2005}.
Such an architecture keeps the physical wiring cost low while supporting high functional
complexity and information-processing capacity~\cite{BullmoreSporns2009,BullmoreSporns2012}.
This fractal-like self-similarity has no natural counterpart in $\mathbb R^N$ with local
coupling, but it is built into $\Qp^N$, where $\Qp$ denotes the field of $p$-adic numbers,
which naturally form a tree-like, self-similar network. Here we
show that this hierarchy, not the dimension of space, decides whether avalanches are
mean-field. We formulate the Buice--Cowan theory on a hierarchical neuron space, the
$p$-adic numbers, and find that the upper critical dimension is $N_c=2\sigma$, where
$\sigma$ measures how fast the coupling decays up the hierarchy. Any $\sigma>0$ is allowed,
so the $3/2$ law holds even for $N=1$ when $\sigma\le1/2$, and exact simulations confirm
the predicted crossover. The mathematical details, including the exact discretization,
the operator-selection and uniqueness results, the heat-kernel estimates, the derivation
of the Ginzburg criterion and the one-loop equations, are worked out in a companion
paper~\cite{companion}, which also contains the further results mentioned below.

\begin{figure}[t]
\centering
\includegraphics[width=\columnwidth]{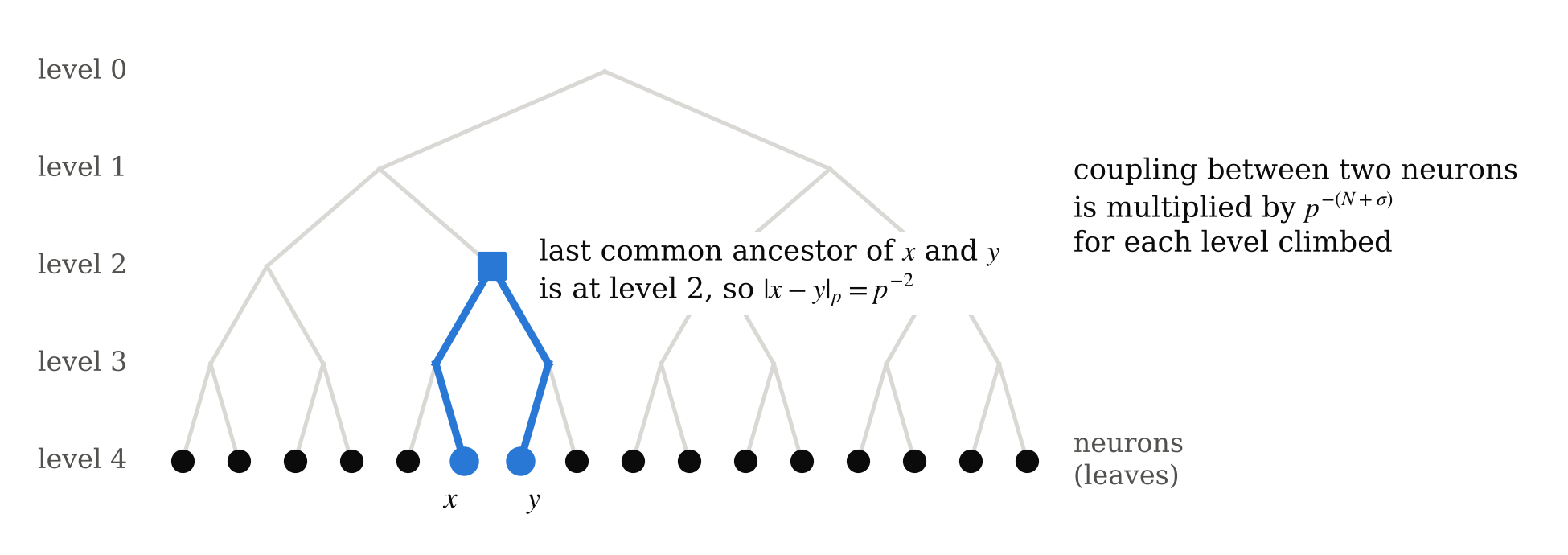}
\caption{Neurons organized in a tree-like structure ($p=2$, depth $4$: levels $0,\dots,4$). The $p$-adic
distance between two neurons is determined by the level of their last common ancestor.
With a kernel $w(\norm{x})\propto\norm{x}^{-N-\sigma}$, the exponent $\sigma$ measures how
fast coupling decays with hierarchical distance.}
\label{fig:tree}
\end{figure}

\begin{figure*}[t]
\centering
\includegraphics[width=0.92\textwidth]{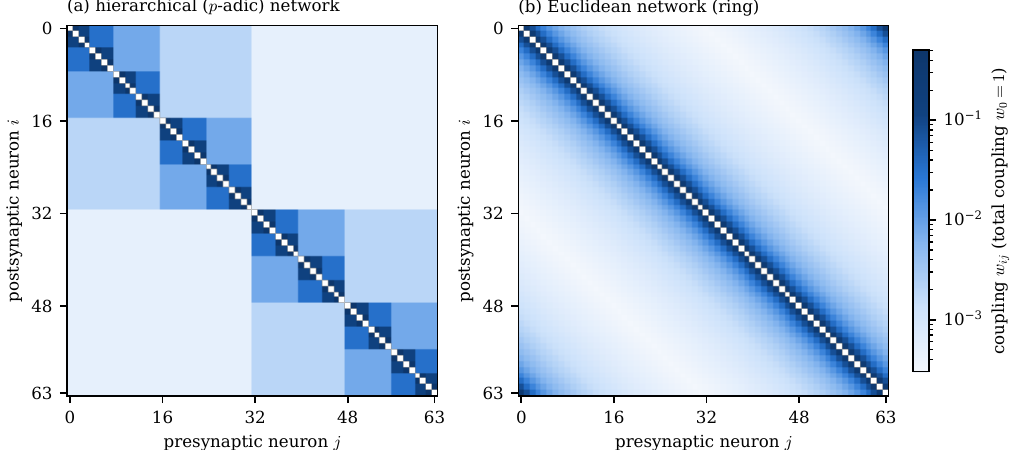}
\caption{Connection matrices $w_{ij}$ of two networks of $64$ neurons with $\sigma=1$
(logarithmic color scale; diagonal excluded; total coupling per neuron normalized to
$1$). (a) Hierarchical network, $p=2$, $N=1$, $l=6$, $w_{ij}\propto|i-j|_2^{-(1+\sigma)}$.
Neurons are ordered along the boundary of the tree; the nested blocks are the modules of
levels $5,4,\dots,1$, and $w$ is constant on each block. (b) Euclidean ring with
$w_{ij}\propto|i-j|^{-(1+\sigma)}$.}
\label{fig:matrices}
\end{figure*}

\paragraph*{$p$-Adic neuron spaces.}
Ultrametric spaces entered physics through the theory of spin glasses: in Parisi's
solution of mean-field models, the pure states are organized in an ultrametric
hierarchy~\cite{Mezard1984}. The $p$-adic numbers are the prototypical ultrametric spaces
on which one can do analysis, and they have been used to formulate replica symmetry
breaking~\cite{ParisiSourlas2000} and relaxation in hierarchical energy
landscapes~\cite{ABKO2002}. Here the hierarchy is not in the space of states but in the
architecture: the neurons themselves are the points of an ultrametric space.
Fix a prime $p$. Every nonzero $p$-adic number has a unique expansion
$x=p^{\gamma}\sum_{j\ge0}x_jp^j$, with digits $x_j\in\{0,\dots,p-1\}$, $x_0\neq0$, and
$|x|_p=p^{-\gamma}$~\cite{VVZ1994,ZunigaLNM2016}. The norm satisfies the
strong triangle inequality $|x+y|_p\le\max\{|x|_p,|y|_p\}$. Consequently two balls are
either disjoint or nested, and the unit ball $\Zp$ is the boundary of a rooted tree in which
each vertex has $p$ children. The distance $|x-y|_p$ records the level of the last common
ancestor of $x$ and $y$ (Fig.~\ref{fig:tree}). In dimension $N$ we use $\Qp^N$ with
$\norm{x}=\max_i|x_i|_p$; it carries a Haar measure $d^Nx$, a Fourier transform and a
theory of pseudodifferential operators~\cite{VVZ1994,ZunigaLNM2016}. A
finite network of depth $l$ has as neurons the $p^{Nl}$ elements of
$G_l^N=(\Zp/p^l\Zp)^N$, the leaves of a finite tree of depth $l$; the internal vertices
are modules (balls). We say that the neurons are organized in a tree-like structure, but a
hierarchical network includes a tree as a particular case. In general, every pair of
neurons is coupled, with a strength $w_{ij}=w(\norm{i-j})$ fixed by the level of their last common
ancestor
[Fig.~\ref{fig:matrices}(a)], and the neuron space has finite dimension $N$, since a ball
of radius $p^k$ has Haar volume $p^{kN}$.

\paragraph*{Model.}
We use the effective spike model of Buice and
Cowan~\cite{BuiceCowan2007}: each effective spike decays at rate $\alpha$, and neuron $i$
emits spikes at rate $f\big(\sum_jw_{ij}n_j\big)$, with $f(0)=0$ (an absorbing state),
$f'(0)>0$ and $f''(0)<0$ (saturation). The absorbing state is an idealization: resting
cortex receives weak input, $f(I)>0$, and is described by a weakly driven state near
criticality, in which input spikes start avalanches; the absorbing state is its limit of
zero input. The Doi--Peliti construction gives the action
\begin{equation}
\begin{aligned}
S[\varphi,\tphi]=\int_{\Qp^N}\!d^Nx\!\int_0^\infty\!dt\,\Big\{&\tphi\,\partial_t\varphi
+\alpha\tphi\varphi\\
&-\tphi\,f\big(w\star(\tphi\varphi+\varphi)\big)\Big\},
\end{aligned}
\label{eq:action}
\end{equation}
with $(w\star g)(x)=\int w(\norm{x-y})g(y)\,d^Ny$; the initial-state term is omitted. Unlike
a lattice approximation of $\mathbb R^N$, the network $G_l^N$ is an \emph{exact}
restriction of the continuum theory: for fields constant on the balls of radius $p^{-l}$,
the discrete action equals Eq.~\eqref{eq:action} for every $l$~\cite{companion}. Setting
$\tphi=0$ in the equations of motion gives the hierarchical Wilson--Cowan equation
$\partial_ta=-\alpha a+f(w\star a)$.

\paragraph*{The connectivity selects the diffusion operator.}
Linearizing around the quiescent state, a mode of momentum $\xi\in\Qp^N$ relaxes at the
rate $\lambda(\xi)=\mu+A_w(\xi)$, exactly, where $\mu=\alpha-f'(0)\hat w(0)$ is the bare
distance to criticality and
\begin{equation}
A_w(\xi)=f'(0)\big(\hat w(0)-\hat w(\xi)\big)
\label{eq:symbol}
\end{equation}
is the symbol of the nonlocal operator
$\boldsymbol W\phi(x)=f'(0)\int[\phi(x)-\phi(x-y)]\,w(\norm{y})\,d^Ny$. Thus the free part
of the theory is $\tphi(\partial_t+\mu+\boldsymbol W)\varphi$, with no approximation.

In $\mathbb R^N$ the same construction gives an operator whose long-wavelength limit is
$-D\nabla^2$ whenever $w$ has a finite second moment: all short-range kernels reduce to the
Laplacian. Over $\Qp^N$ there is no such reduction, because locally constant functions have
no Taylor expansion. Each connectivity kernel selects its own heat equation
$\partial_tu+\boldsymbol Wu=0$, and the infrared physics depends on how $A_w(\xi)$
behaves as $\norm{\xi}\to0$. The natural hierarchical kernel attenuates the coupling by
the same factor at every level,
\begin{equation}
\frac{w(p^{j+1})}{w(p^{j})}=p^{-(N+\sigma)},\qquad\text{i.e.}\quad
w(\norm{x})=C_w\norm{x}^{-N-\sigma},
\label{eq:perlevel}
\end{equation}
for $\norm{x}$ larger than some $p^r$. Then, exactly for $\norm{\xi}\le p^{-r}$,
\begin{equation}
A_w(\xi)=\kappa_W\norm{\xi}^{\sigma},\qquad
\kappa_W=f'(0)\,C_w\,\frac{1-p^{-\sigma-N}}{p^{\sigma}-1},
\label{eq:tail}
\end{equation}
which is the symbol of $\kappa_W\Dsig$, with $\Dsig$ the Taibleson--Vladimirov
operator~\cite{VVZ1994,ZunigaLNM2016}. Conversely, $\Dsig$ is the only isotropic
pseudodifferential operator on $\Qp^N$ that is homogeneous of degree
$\sigma$~\cite{companion}. Self-similar hierarchical connectivity therefore fixes the
diffusion operator up to one constant. Its heat kernel has power-law tails in
hierarchical distance,
$Z_t(x)\asymp\kappa_Wt\,[(\kappa_Wt)^{1/\sigma}+\norm{x}]^{-N-\sigma}$, with constants that
depend only on $p$, $N$ and $\sigma$~\cite{BGPW2014,companion}.

\paragraph*{Ultrametric directed percolation and $N_c=2\sigma$.}
Expanding $f$ to second order, shifting the response field and rescaling as in
Ref.~\cite{BuiceCowan2007}, the relevant part of Eq.~\eqref{eq:action} becomes
\begin{equation}
S_{\rm eff}=\int d^Nx\,dt\,\Big\{\tpsi\big(\partial_t+\mu+\boldsymbol W\big)\psi
+\frac g2\big(\tpsi\psi^2-\tpsi^2\psi\big)\Big\},
\label{eq:RFT}
\end{equation}
with $g=[2f'(0)|f''(0)|\hat w(0)^3]^{1/2}$. This is Reggeon field theory with $-D\nabla^2$
replaced by $\boldsymbol W$, and it keeps the rapidity-reversal symmetry
$\psi(x,t)\leftrightarrow-\tpsi(x,-t)$, since $\boldsymbol W$ is self-adjoint. Power counting
with $A_w\simeq\kappa_W\norm{\xi}^\sigma$ gives $z=\sigma$ at tree level,
$[\psi]=[\tpsi]=N/2$, and
\begin{equation}
[g]=\sigma-\frac N2,\qquad N_c=2\sigma .
\label{eq:Nc}
\end{equation}
Above $N_c$ the mean-field exponents
\begin{equation}
\beta=1,\quad\nu_\parallel=1,\quad\nu_\perp=\frac1\sigma,\quad z=\sigma,\quad\tau_S=\frac32
\label{eq:MF}
\end{equation}
are exact; at $N=N_c$ they acquire logarithmic corrections. Quartic and higher vertices
are irrelevant near $N_c$, and a correction $O(\norm{\xi}^{\sigma+\delta})$ to the symbol
does not change the power counting.

The comparison with Euclidean long-range DP is instructive. With L\'evy-flight
spreading, $w(x)\sim|x|^{-N-\sigma}$ in $\mathbb R^N$, one also finds $N_c=2\sigma$, but only
for $\sigma<2$: for $\sigma\ge2$ the second moment is finite, $\nabla^2$ takes over, and
$N_c=4$~\cite{Janssen1999,HinrichsenHoward1999}. Over $\Qp^N$ no local operator exists, so
\emph{every} $\sigma>0$ defines a distinct universality class, and $N_c=2\sigma$ ranges over
$(0,\infty)$. In particular, for $N=1$ the mean-field avalanche law $\tau_S=3/2$ holds
whenever $\sigma\le1/2$, while for $N=2$ it holds whenever $\sigma\le1$. The situation is
analogous to Dyson's hierarchical model~\cite{Dyson1969,BleherSinai1973}, which also has a
finite upper critical dimension set by the decay of the coupling. It is not that of DP on a
Cayley tree, which is mean-field for any coupling because the number of sites grows
exponentially with graph distance.

\paragraph*{Ginzburg criterion.}
The one-loop correction to the inverse propagator is small compared with $\mu$ if
\begin{equation}
\frac{g^2}{4}\,I_2(\mu)\ll1,\qquad
I_2(\mu)=\int_{\norm{\xi}\le p^{\Lambda}}\frac{d^N\xi}{[\mu+A_w(\xi)]^2}.
\label{eq:I2def}
\end{equation}
$I_2(0)$ is finite if and only if $N>2\sigma$, consistent with Eq.~\eqref{eq:Nc}. For
$N<2\sigma$ (see Ref.~\cite{companion} for the proof), summing over the spheres $\norm{\xi}=p^k$ gives, as $\mu\to0^+$,
\begin{equation}
I_2(\mu)\simeq(1-p^{-N})\,\kappa_W^{-N/\sigma}\mu^{N/\sigma-2}\,\Phi(s),
\label{eq:I2}
\end{equation}
where $\mu/\kappa_W=p^{\sigma s}$ and
\begin{equation}
\Phi(s)=\sum_{k\in\mathbb Z}\frac{p^{(k-s)N}}{\big(1+p^{\sigma(k-s)}\big)^2}
\label{eq:Phi}
\end{equation}
is periodic in $s$ with period $1$. With the ultrametric diffusion length
$\ell_W=(\kappa_W/\mu)^{1/\sigma}$, the criterion reads
\begin{equation}
\frac{1-p^{-N}}{4}\,\frac{g^2}{\kappa_W^2}\,\ell_W^{\,2\sigma-N}\,\Phi(s)\ll1 .
\label{eq:ginzburg}
\end{equation}
The left-hand side is the relative size of the fluctuation correction to the response of
the network, i.e., to the mean avalanche size. The coupling $g^2\propto f'(0)|f''(0)|$
measures the spike-count noise that survives in the dynamics of the mean; it comes from
branching and saturation, and vanishes for a linear $f$, for which mean-field theory is
exact. The factor $\ell_W^{\,2\sigma-N}$ measures how much of this noise is averaged out: a
perturbation lives a time $1/\mu=\ell_W^{\sigma}/\kappa_W$ and spreads over a ball of radius
$\ell_W$, which contains a number of neurons proportional to $\ell_W^{N}$. As $\mu\to0$, this
ball climbs the hierarchy. If $N>2\sigma$, the number of neurons it contains grows faster
than the noise accumulates, and mean-field theory improves; if $N<2\sigma$, the coupling
decays too fast up the hierarchy, the effective number of independent inputs does not keep
up, and fluctuations dominate. Three consequences follow. First, on the infinite network and for $N<2\sigma$, mean-field
theory fails sufficiently close to criticality, whatever the value of $g$. Second, on a
finite network of diameter $L$, the left-hand side is largest at $\ell_W\approx L$, so
mean-field theory fails inside the scaling window $p^{-l}\ll\ell_W\ll L$ only if the
Ginzburg number $\mathrm{Gi}=g^2L^{2\sigma-N}/\kappa_W^2$ is of order one or larger; otherwise
finite-size rounding sets in first. Third, the factor $\Phi(s)$ makes the criterion periodic
in $\ln\mu$ with period $\sigma\ln p$: the hierarchy is invariant only under the discrete
dilations $x\mapsto p^{-1}x$, $t\mapsto p^\sigma t$, and the theory inherits this discrete
scale invariance. A Euclidean theory with a continuous scaling symmetry cannot produce
such log-periodic corrections. Their size is, however, controlled by the Poisson
summation of Eq.~\eqref{eq:Phi}: the first harmonic has amplitude of order
$|\omega_1|e^{-\pi|\omega_1|}$, with $\omega_1=2\pi/(\sigma\ln p)$. For $N=\sigma=1$ the relative
amplitude is about $5\times10^{-11}$ for $p=2$, $2\times10^{-3}$ for $p=7$, $0.07$ for $p=31$
and $0.24$ for $p=101$. Log-periodicity is therefore a qualitative signature of a discrete
hierarchy, measurable only when $\sigma\ln p$ is large.

\begin{figure*}[t]
\centering
\includegraphics[width=0.92\textwidth]{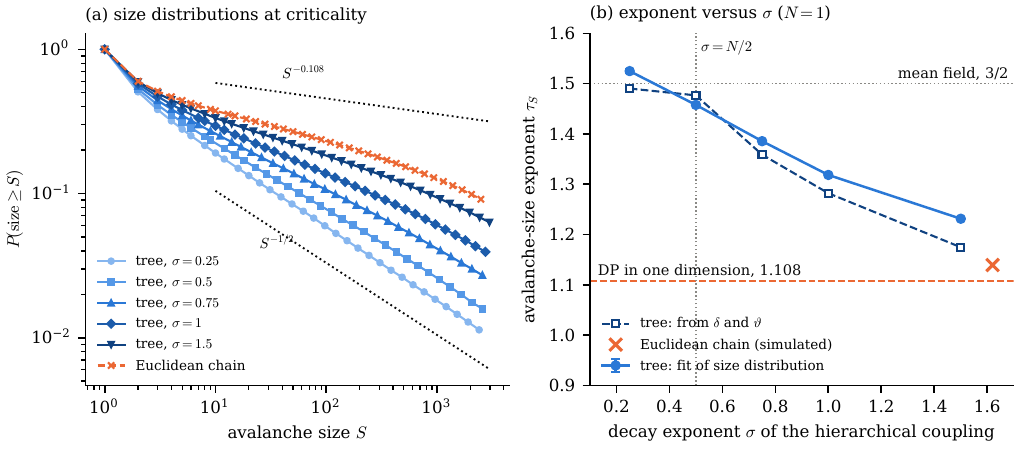}
\caption{Avalanches at the critical point. (a) Probability that an avalanche has size at
least $S$, for the hierarchical network ($p=2$, $N=1$, $l=40$) with five values of $\sigma$,
and for a Euclidean chain with nearest-neighbor coupling. Dotted lines: the mean-field law
$S^{-1/2}$ ($\tau_S=3/2$) and the DP law $S^{-0.108}$ in one dimension. (b) $\tau_S$ against
$\sigma$. Filled circles: maximum-likelihood fit of the size distribution
($10\le S\le2000$). Open squares: $\tau_S=1+\delta/(1+\vartheta+\delta)$ from the measured
survival and growth exponents. The vertical line marks $\sigma=N/2$, where $N=N_c$.}
\label{fig:avalanches}
\end{figure*}

\paragraph*{Simulations.}
To test the prediction beyond power counting, we simulated the effective spike model
directly on the tree $G_l$ with $p=2$, $N=1$ and $l=40$, with
$w_{ij}\propto2^{-k(1+\sigma)}$ for two neurons whose smallest common module has $2^k$ neurons
[i.e., $\norm{i-j}=2^{k-l}$; this is Eq.~\eqref{eq:perlevel} with $N=1$]. The simulations
follow standard approaches of nonequilibrium statistical physics; the procedure is
described in Methods.

\begin{table}[t]
\caption{Critical point $b_c$ (mean-field value $1$) and exponents: survival
$P(t)\propto t^{-\delta}$, growth $\mathcal N(t)\propto t^{\vartheta}$, and $\tau_S$ from the
fit and from scaling. In parentheses, the exact values of $(\delta,\vartheta,\tau_S)$ for the
two controls: the critical branching process, and DP in one dimension. The controls show
a fit bias of about $+0.03$.}
\label{tab:sim}
\begin{ruledtabular}
\begin{tabular}{lccccc}
 & $b_c$ & $\delta$ & $\vartheta$ & $\tau_S^{\rm fit}$ & $\tau_S^{\rm scal}$\\
\hline
Branching ($1$, $0$, $3/2$) & 1.000 & 1.02 & $-0.06$ & 1.53 & 1.52\\
$\sigma=0.25$ & 1.106 & 0.99 & $+0.03$ & 1.52 & 1.49\\
$\sigma=0.5$ & 1.371 & 0.90 & $-0.01$ & 1.46 & 1.48\\
$\sigma=0.75$ & 1.739 & 0.65 & $+0.16$ & 1.39 & 1.36\\
$\sigma=1$ & 2.169 & 0.49 & $+0.24$ & 1.32 & 1.28\\
$\sigma=1.5$ & 3.157 & 0.28 & $+0.32$ & 1.23 & 1.17\\
Chain ($0.159$, $0.314$, $1.108$) & 4.848 & 0.16 & $+0.31$ & 1.14 & 1.11\\
\end{tabular}
\end{ruledtabular}
\end{table}

Table~\ref{tab:sim} and Fig.~\ref{fig:avalanches} show three results. (i) The controls are
reproduced: the chain gives the one-dimensional DP exponents, and the branching process
gives $\delta\simeq1$, $\vartheta\simeq0$. In both cases the fitted $\tau_S$ exceeds the exact
value by about $0.03$, which we take as the systematic error of the fit. (ii) For
$\sigma=0.25<N/2$ the exponents are those of the branching process, $\tau_S\simeq3/2$, and at
$\sigma=N/2=0.5$ they are close to it, with deviations of the size expected from logarithmic
corrections at $N=N_c$. (iii) For $\sigma>N/2$, $\tau_S$ decreases continuously, to about
$1.2$ at $\sigma=1.5$, approaching the Euclidean value $1.11$. The critical point also moves
away from its mean-field value $b_c=1$ as $\sigma$ grows.

The mechanism is visible in the dynamics. On the chain a spike excites only its nearest
neighbors, so activity spreads as a compact patch, new spikes keep landing on neurons that
already receive input, and the saturation of $f$ penalizes this crowding. On the tree with
small $\sigma$ the coupling decays slowly with hierarchical distance, most new spikes land far
from the active neurons, crowding is rare, and the avalanches are those of a branching
process. As $\sigma$ grows, activity stays inside small modules and crowding returns.

\paragraph*{Discussion.}
In the Euclidean theory with short-range coupling, the dimension of space decides whether
neural avalanches are mean-field, and the Beggs--Plenz exponent requires $N\ge4$. In the
hierarchical theory the decay of the coupling up the hierarchy decides: $\tau_S=3/2$ for
$\sigma\le N/2$, and non-mean-field exponents for $\sigma>N/2$. This yields a falsifiable
prediction. Measure how connection strength decays with hierarchical distance, for example
from connectome data organized by modules, and extract $\sigma$ from
Eq.~\eqref{eq:perlevel}. The theory then predicts whether avalanche statistics are
mean-field or anomalous. A cortical sheet with $N=2$ needs only $\sigma\le1$.

The theory applies equally to artificial networks whose dynamics is a hierarchical
Wilson--Cowan equation, such as $p$-adic Wilson--Cowan and cellular neural
networks~\cite{ZZ2023WC,ZZ2023CNN}.

Several questions remain open. The critical exponents for $N<2\sigma$ require the
$p$-adic renormalization group~\cite{LernerMissarov1989,AbdesselamChandraGuadagni2013};
the simulations show that they interpolate between the mean-field and the Euclidean DP
values, but their values are not yet known analytically. The simulations cover one tree
($p=2$, $N=1$) and one saturation scale. Finally, hierarchical networks with quenched
heterogeneity show extended critical (Griffiths) regions~\cite{MorettiMunoz2013}; the
translation-invariant $p$-adic model has a sharp transition, and level-dependent disorder
would interpolate between the two pictures. The companion paper~\cite{companion} gives the
full construction, the loop-corrected hierarchical Wilson--Cowan equations, the small-world
property of these networks, and the phase transition and self-organization.

\paragraph*{Methods.}
The simulations use standard methods for absorbing-state transitions~\cite{Hinrichsen2000}.
\emph{Dynamics.} The network has $2^{40}$ neurons, so that the boundary is essentially never
reached; there is no self-coupling, the total coupling is $w_0=1$, $\alpha=1$, and
$f(s)=f_{\max}(1-e^{-s/s_0})$ with $s_0=0.25$. The continuous-time dynamics is simulated
exactly, by an event-driven (Gillespie-type) algorithm~\cite{Gillespie1976} combined with
thinning~\cite{LewisShedler1979}: since $f(s)\le f'(0)s$, each spike attempts to trigger a
spike at rate $b=f'(0)w_0/\alpha$ at a neuron $i$ chosen with probability $w_{ij}/w_0$, and the
attempt is accepted with probability $f(s_i)/[f'(0)s_i]$, so that the saturation is treated
without approximation. \emph{Spreading from a seed.} Each run starts from a single
spike, as in the spreading method of Grassberger and de la Torre~\cite{GrassbergerTorre1979}.
For each $\sigma$ the critical value $b_c$ was located by bisection, requiring the mean
number of spikes $\mathcal N(t)$ to be a pure power law up to $t=10^3$. At $b_c$ we
generated $2\times10^5$ runs and measured the survival probability $P(t)\propto t^{-\delta}$
and $\mathcal N(t)\propto t^{\vartheta}$ by least-squares fits on logarithmic scales for
$30\le t\le10^3$. \emph{Avalanche sizes.} The size $S$ is the total number of spikes. The
exponent $\tau_S$ was obtained by maximum likelihood for a power law truncated to
$10\le S\le2000$~\cite{Clauset2009}, with bootstrap errors, and independently from the
scaling relation $\tau_S=1+\delta/(1+\vartheta+\delta)$~\cite{Munoz1999}. \emph{Controls.} The
same program was run on two cases with known exponents, a pure branching process and a
Euclidean chain with nearest-neighbor coupling (one-dimensional DP); they calibrate the
systematic bias of the fit. The simulation code, the data and the scripts that produce the figures are available at
\url{https://wazuniga.github.io/padic-BC/}. We used Claude (Anthropic; model \mbox{claude-opus-5-5}) to help implement the numerical
simulation code for some well-known techniques in statistical physics
Refs.~\cite{Hinrichsen2000,Gillespie1976,LewisShedler1979,GrassbergerTorre1979}. The control
was carried out as discussed before.

\paragraph*{Acknowledgments.}
During the preparation of this work, the author used Claude (Anthropic; model
\mbox{claude-opus-5-5}) to improve the clarity and readability of the language of select
sections. After using this tool, the authors reviewed and edited the content as needed and
took full responsibility for the publication. The author was partially supported by the
Debnath Endowment.


\begin{thebibliography}{99}

\bibitem{BeggsPlenz2003}
J.~M. Beggs and D. Plenz, J. Neurosci. \textbf{23}, 11167 (2003).

\bibitem{BuiceCowan2007}
M.~A. Buice and J.~D. Cowan, Phys. Rev. E \textbf{75}, 051919 (2007).

\bibitem{Munoz1999}
M.~A. Mu\~noz, R. Dickman, A. Vespignani, and S. Zapperi, Phys. Rev. E \textbf{59},
6175 (1999).

\bibitem{Sporns2006}
O. Sporns, BioSystems \textbf{85}, 55 (2006).

\bibitem{CostaSporns2005}
L.~da F. Costa and O. Sporns, Eur. Phys. J. B \textbf{48}, 555 (2005).

\bibitem{SpornsTononiKotter2005}
O. Sporns, G. Tononi, and R. K\"otter, PLoS Comput. Biol. \textbf{1}, e42 (2005).

\bibitem{BullmoreSporns2009}
E. Bullmore and O. Sporns, Nat. Rev. Neurosci. \textbf{10}, 186 (2009).

\bibitem{BullmoreSporns2012}
E. Bullmore and O. Sporns, Nat. Rev. Neurosci. \textbf{13}, 336 (2012).

\bibitem{companion}
W.~A. Z\'u\~niga-Galindo, $p$-Adic statistical field theory of fluctuation effects in
hierarchical neural networks. In preparation; available at arXiv.org.

\bibitem{Mezard1984}
M. M\'ezard, G. Parisi, N. Sourlas, G. Toulouse, and M. Virasoro, J. Phys. (Paris)
\textbf{45}, 843 (1984).

\bibitem{ParisiSourlas2000}
G. Parisi and N. Sourlas, Eur. Phys. J. B \textbf{14}, 535 (2000).

\bibitem{ABKO2002}
V.~A. Avetisov, A.~H. Bikulov, S.~V. Kozyrev, and V.~A. Osipov, J. Phys. A
\textbf{35}, 177 (2002).

\bibitem{VVZ1994}
V.~S. Vladimirov, I.~V. Volovich, and E.~I. Zelenov, \emph{$p$-Adic Analysis and
Mathematical Physics} (World Scientific, Singapore, 1994).

\bibitem{ZunigaLNM2016}
W.~A. Z\'u\~niga-Galindo, \emph{Pseudodifferential Equations over Non-Archimedean
Spaces}, Lecture Notes in Mathematics Vol.~2174 (Springer, Cham, 2016).

\bibitem{BGPW2014}
A. Bendikov, A. Grigor'yan, Ch. Pittet, and W. Woess, Russ. Math. Surveys \textbf{69},
589 (2014) [arXiv:1304.6271].

\bibitem{Janssen1999}
H.~K. Janssen, K. Oerding, F. van Wijland, and H.~J. Hilhorst, Eur. Phys. J. B \textbf{7}, 137 (1999).

\bibitem{HinrichsenHoward1999}
H. Hinrichsen and M. Howard, Eur. Phys. J. B \textbf{7}, 635 (1999).

\bibitem{Dyson1969}
F.~J. Dyson, Commun. Math. Phys. \textbf{12}, 91 (1969).

\bibitem{BleherSinai1973}
P.~M. Bleher and Ya.~G. Sinai, Commun. Math. Phys. \textbf{33}, 23 (1973).

\bibitem{ZZ2023WC}
W.~A. Z\'u\~niga-Galindo and B.~A. Zambrano-Luna, Entropy \textbf{25}, 949 (2023).

\bibitem{ZZ2023CNN}
B.~A. Zambrano-Luna and W.~A. Z\'u\~niga-Galindo, $p$-adic cellular neural networks,
J. Nonlinear Math. Phys. \textbf{30}, 34--70 (2023), doi:10.1007/s44198-022-00071-8.

\bibitem{LernerMissarov1989}
\'E.~Yu. Lerner and M.~D. Missarov, Theor. Math. Phys. \textbf{78}, 177 (1989).

\bibitem{AbdesselamChandraGuadagni2013}
A. Abdesselam, A. Chandra, and G. Guadagni, Rigorous quantum field theory functional
integrals over the $p$-adics I: anomalous dimensions, arXiv:1302.5971.

\bibitem{MorettiMunoz2013}
P. Moretti and M.~A. Mu\~noz, Nat. Commun. \textbf{4}, 2521 (2013).

\bibitem{Hinrichsen2000}
H. Hinrichsen, Adv. Phys. \textbf{49}, 815 (2000).

\bibitem{Gillespie1976}
D.~T. Gillespie, J. Comput. Phys. \textbf{22}, 403 (1976).

\bibitem{LewisShedler1979}
P.~A.~W. Lewis and G.~S. Shedler, Nav. Res. Logist. Q. \textbf{26}, 403 (1979).

\bibitem{GrassbergerTorre1979}
P. Grassberger and A. de la Torre, Ann. Phys. (N.Y.) \textbf{122}, 373 (1979).

\bibitem{Clauset2009}
A. Clauset, C.~R. Shalizi, and M.~E.~J. Newman, SIAM Rev. \textbf{51}, 661 (2009).

\end{thebibliography}
\end{document}